\documentclass[journal, onecolumn, draftclsnofoot]{IEEEtran}
\usepackage{amsmath,amsfonts}
\usepackage{algorithmic}
\usepackage{algorithm}
\usepackage{array}
\usepackage[caption=false,font=normalsize,labelfont=sf,textfont=sf]{subfig}
\usepackage{textcomp}
\usepackage{stfloats}
\usepackage{url}
\usepackage{verbatim}
\usepackage{graphicx}
\usepackage{booktabs}
\usepackage{cite}
\usepackage{amsmath,amsthm}
\usepackage{amssymb}
\newtheorem{theorem}{Theorem}
\newtheorem{lemma}{Lemma}
\newtheorem{corollary}{Corollary}
\newtheorem{proposition}{Proposition}
\newtheorem{definition}{Definition}
\newtheorem{remark}{Remark}
\begin{document}

\title{Galois EA-CSS Construction and Entanglement Control via Index Selection for Fixed Classical Code Pairs}

\author{Mengqing Yang,~Tongjiang Yan~
}




\maketitle

\begin{abstract}
Entanglement-assisted quantum code constructions and their entanglement requirements are closely related to the pairing structure of the underlying classical codes. For a prescribed ordered pair of classical linear codes over a finite field, this paper presents an EA-CSS stabilizer construction at arbitrary Galois indices and investigates how index selection controls entanglement consumption while keeping the code pair and its coordinate arrangement fixed. Frobenius twisting of the second classical code realizes the Galois pairing through the standard symplectic form. An explicit stabilizer completion determines the number of encoded logical qudits and yields an exact expression for the minimum distance. At each index, the minimum entanglement cost among \(\mathbb{F}_q\)-linear stabilizer completions preserving the prescribed label spaces equals the Galois cross Gram rank and is equivalently determined by the Galois relative hull dimension. The Euclidean pairing and, over finite fields of square order, the Hermitian pairing are recovered as special cases. Across indices, entanglement cost and logical dimension vary by the same amount, while the Frobenius periods of the input codes can reduce the index search. Intersection formulas determining the exact entanglement costs are further established for fixed GRS and BCH code pairs. The GRS constructions yield entanglement-assisted quantum MDS codes with the same exact minimum distance across Galois indices, whereas the BCH constructions provide a minimum distance lower bound independent of the index. Explicit constructions demonstrate that suitable Galois-index choices can achieve entanglement costs strictly below the Euclidean and Hermitian levels, with zero-entanglement realizations for certain fixed GRS pairs.
\end{abstract}


\section{Introduction}
\IEEEPARstart{Q}{uantum} error-correcting codes protect quantum information against noise by encoding logical quantum states redundantly across physical qubits or qudits \cite{bib1,bib2,bib3,bib4,bib5,bib6,bib7}. Stabilizer constructions connect quantum codes with classical coding theory, subject to suitable self-orthogonality conditions. The entanglement-assisted framework relaxes these restrictions by using pre-shared entanglement to compensate for noncommuting stabilizer generators \cite{bib8,bib9}. Rank characterizations of the minimum entanglement requirement \cite{bib10} and their extension to arbitrary finite fields \cite{bib11} provide an algebraic basis for quantifying the required resources.

Entanglement consumption is closely related to the orthogonality structure of the underlying classical codes. Guenda, Jitman, and Gulliver \cite{bib12} established the connection between the hull dimension and entanglement requirements, providing a basis for regulating quantum resources through the orthogonal structure of classical codes. The linear complementary dual (LCD) codes proposed by Massey correspond to the case of a trivial hull \cite{13}; Carlet et al. \cite{14} proved that, under appropriate alphabet conditions, linear codes can be transformed into an LCD representation via equivalence transformations. For MDS codes, Luo, Cao, and Chen \cite{15} investigated constructions with a prescribed hull dimension and their quantum applications, and Fang et al. \cite{16} further utilized GRS codes and extended GRS codes to construct MDS codes with tunable Euclidean or Hermitian hull dimensions. Complementing hull-based constructions, Luo et al.\cite{bib22} developed propagation rules for varying entanglement consumption and analyzed their effects on error-correction performance. These results show that the orthogonality structure is not determined solely by the classical length, dimension, and minimum distance. Code design and suitable equivalence transformations therefore provide ways to control hull dimensions and the associated entanglement requirements.

The explicit determination of entanglement requirements also relies on the specific algebraic structure of classical codes. Sangwisut et al.\cite{23} utilized generator polynomials to investigate the Euclidean and Hermitian hulls of cyclic and negacyclic codes; Galindo et al.\cite{24} established dual-code constructions for asymmetric entanglement-assisted quantum codes and computed explicit parameters for those derived from BCH codes. Reed-Solomon codes \cite{25} and their generalized forms provide a standard MDS setting for explicit quantum code constructions. Recently, Campion and San-Jos\'e\cite{27}, focusing on specific families of GRS codes, transformed the computation of the Hermitian hull dimension into lattice point counting, thereby obtaining explicit expressions for entanglement consumption and new parameters for EAQMDS codes. These studies illustrate how the evaluation structure of GRS codes and the defining-set structure of cyclic codes make hull dimensions and entanglement requirements amenable to explicit algebraic or combinatorial computation.

The Euclidean and Hermitian notions of orthogonality underlying these constructions can be described within the Galois inner-product framework introduced by Fan and Zhang \cite{17}. Over \(\mathbb F_{p^m}\), the Galois inner product reduces to the Euclidean inner product at \(\ell=0\) and to the Hermitian inner product at \(\ell=m/2\) when \(m\) is even. Liu and Pan \cite{18} derived a generator-matrix formula for the Galois hull dimension, and Liu, Fan, and Liu \cite{19} studied Galois LCD codes. In quantum code construction, Liu, Yu, and Hu \cite{liu2019new} used \(k\)-Galois duality to construct entanglement-assisted quantum codes from a single classical code. Similarly, Cao \cite{20} constructed further quantum code families from GRS and extended GRS codes with prescribed Galois hull dimensions, also starting from a single classical code. For pairs of classical codes, Anderson et al. \cite{21} introduced the Galois relative hull and showed that it can be computed as the Euclidean relative hull of a Frobenius-twisted pair. Their quantum code applications employed the existing Euclidean EA-CSS construction and varied entanglement consumption by replacing one constituent code with a monomially equivalent code.

In light of these studies, for two independently specified classical linear codes of the same length over \(\mathbb F_q\), how can their pairing at a prescribed Galois index be explicitly realized within the standard EA-CSS stabilizer framework, and how can the resulting quantum code parameters be determined? When the ordered code pair and its coordinate arrangement are fixed, what entanglement costs are attainable by varying only the Galois index, and how many logical qudits are encoded at each cost? Under a minimum-distance lower bound independent of the index, when can index selection yield an entanglement cost strictly below the Euclidean level and, over finite fields of square order, below the Hermitian level as well?

To address the construction question, this paper presents an \(\mathbb F_q\)-linear EA-CSS stabilizer construction for a prescribed ordered pair of classical linear codes at a given Galois index. A Frobenius twist of the second code realizes the Galois pairing in the standard symplectic space. An explicit Bob-side stabilizer completion determines the number of encoded logical qudits and yields an exact expression for the minimum distance. Among \(\mathbb F_q\)-linear stabilizer completions that preserve the prescribed label spaces, the minimum entanglement cost is proved to equal the rank of the Galois cross Gram matrix and can equivalently be expressed in terms of the Galois relative hull dimension.

Building on this construction, the original ordered code pair and its coordinate arrangement are held fixed, and only the Galois index is varied to examine the effect of index selection on entanglement requirements under fixed classical inputs. The entanglement cost and the number of encoded logical qudits change by the same amount: each entangled pair saved is accompanied by one fewer encoded logical qudit. Furthermore, the Frobenius periods of the input codes can reduce the number of Galois indices that need to be examined.

As concrete applications of this framework, explicit EA-CSS constructions are developed from fixed GRS and BCH code pairs. In the GRS setting, modular intersection counts determine the exact entanglement cost, with a closed form expression obtained for a specified shift. For each fixed GRS code pair considered, the resulting EAQMDS codes have the same exact minimum distance across all Galois indices. Defining-set intersections provide the corresponding exact entanglement formula for fixed cyclic code pairs. Specialization to BCH codes allows these intersection counts to be evaluated as weighted overlaps of cyclotomic cosets, while the BCH bound supplies a minimum-distance lower bound independent of the index. Explicit examples demonstrate entanglement costs strictly below both the Euclidean and Hermitian levels, including zero-entanglement realizations for certain fixed GRS code pairs, and show that the set of attainable entanglement costs may contain gaps.

The remainder of this paper is organized as follows. Section II introduces the notation and preliminaries. Section III develops the Galois EA-CSS framework and establishes the resource relations and index-reduction results for fixed classical code pairs. Sections IV and V specialize the framework to GRS and BCH code pairs, respectively, providing explicit formulas and comparisons across Galois indices. Section VI concludes the paper and discusses directions for further research.

\section{Preliminaries}
\subsection{Finite fields and Frobenius twists}
Let $\mathbb{F}_q$ denote a finite field of order $q = p^m$, where $p$ is a prime and $m$ is a positive integer. For any integer $\ell$ with $0\le\ell<m$, the $\ell$-th Frobenius automorphism is defined by
$$
\sigma_\ell(a)=a^{p^\ell},\quad a\in\mathbb{F}_q.
$$
For a vector or matrix $A=(a_{ij})$, denote by $A^{[\ell]}$ its entrywise Frobenius twist, i.e.,
$$
\big(A^{[\ell]}\big)_{ij}=a_{ij}^{p^\ell}.
$$

Since $a^{p^m}=a$ for all $a\in\mathbb{F}_q$, it follows that $\sigma_{m-\ell}$ is precisely the inverse automorphism of $\sigma_\ell$.

For vectors $\textbf{x}=(x_1,x_2,\dots,x_n)$ and $\textbf{y}=(y_1,y_2,\dots,y_n)\in\mathbb{F}_q^n$, the $\ell$-th Galois inner product is defined as
$$
\langle \textbf{x,y}\rangle_\ell=\sum_{i=1}^n x_i y_i^{p^\ell}=\textbf{x}\big(\textbf{y}^{[\ell]}\big)^T.
$$
This paper adopts the convention of the right Galois dual:
$$
C^{\perp_\ell}=\big\{\textbf{y}\in\mathbb{F}_q^n\mid \langle \textbf{x,y}\rangle_\ell=0,\;\forall \textbf{x}\in C\big\}.
$$

Under the right-dual convention, \(y \in C^{\perp \ell}\) if and only if \(y^{[\ell]} \in C^\perp\). Consequently,
\begin{equation}
C^{\perp \ell} = (C^\perp)^{[m-\ell]} = (C^{[m-\ell]})^\perp.
\label{(1)}
\end{equation}
Frobenius twisting preserves dimension and Hamming weight. Hence,
\[
\dim C^{\perp \ell} = n - \dim C, \qquad d(C^{\perp \ell}) = d(C^\perp).
\]
Throughout the paper, Frobenius indices are interpreted modulo \(m\).

\begin{definition}($\ell$-th Galois Adjoint Matrix)
 Let $\mathbb{F}_q$ be a finite field with $q=p^m$. For any matrix $A\in\mathbb{F}_q^{n\times k}$, its $\ell$-th Galois adjoint matrix is defined as
$$
A^{\dagger_\ell}=\big(A^{[m-\ell]}\big)^T.
$$
\end{definition}
The notation \(A^{\dagger_\ell}\) denotes the finite-field Galois adjoint. 

\begin{definition} ($\ell$-th Galois Hull)
 Let $C\subseteq\mathbb{F}_q^n$ be an $[n,r]_q$ linear code, and let $C^{\perp_\ell}$ denote its $\ell$-th right Galois dual. The $\ell$-th Galois hull of $C$ is defined as
$$
\operatorname{Hull}_\ell(C)=C\cap C^{\perp_\ell},
$$
and we denote $h_\ell(C)=\dim\operatorname{Hull}_\ell(C)$. 
\end{definition}
The following standard rank identity is recalled under the right-dual convention; see \cite{18}, \cite{liu2019new}.

\begin{lemma}
Let \(G\) be a full-row-rank generator matrix of $C$. The rank of the Galois Gram matrix satisfies
$$
\operatorname{rank}\big(GG^{\dagger_\ell}\big)=r-h_\ell(C).
$$
\end{lemma}
\begin{remark}
\begin{enumerate}
\item If $\ell=0$, then $A^{\dagger_0}=A^T$, which recovers the usual Euclidean transpose;
\item  If $m$ is even and $\ell=m/2$, then $A^{\dagger_{m/2}}=\big(A^{[m-\ell]}\big)^T$, which coincides with the Hermitian conjugate transpose.
\end{enumerate}
\end{remark}

\subsection{Generalized Pauli operators and symplectic representation}
Let $q=p^m$. The Hilbert space of a single $q$-dimensional quantum subsystem is denoted by
$$
\mathcal{H}_q=\operatorname{span}_{\mathbb{C}}\{|a\rangle:a\in\mathbb{F}_q\}.
$$
Let
$$
\omega=e^{2\pi i/p},
$$
and let
$$
\operatorname{Tr}_{q/p}:\mathbb{F}_q\to\mathbb{F}_p
$$
denote the trace map of the finite field. For arbitrary $a,b\in\mathbb{F}_q$, the generalized Pauli operators are defined as
$$
X(a)|x\rangle=|x+a\rangle,\qquad Z(b)|x\rangle=\omega^{\operatorname{Tr}_{q/p}(bx)}|x\rangle.
$$

For $\mathbf{x}=(x_1,x_2,\dots,x_n)$ and $\mathbf{z}=(z_1,z_2,\dots,z_n)\in\mathbb{F}_q^n$, let
$$
X(\mathbf{x})=\bigotimes_{i=1}^n X(x_i),\qquad Z(\mathbf{z})=\bigotimes_{i=1}^n Z(z_i),
$$
and define
$$
E(\mathbf{x},\mathbf{z})=X(\mathbf{x})Z(\mathbf{z}).
$$
From the generalized Pauli commutation relation, we have
$$
E(\mathbf{x},\mathbf{z})E(\mathbf{x}',\mathbf{z}')=\omega^{\operatorname{Tr}_{q/p}\left(\mathbf{z}\mathbf{x}'^T-\mathbf{x}\mathbf{z}'^T\right)}E(\mathbf{x}',\mathbf{z}')E(\mathbf{x},\mathbf{z}).
$$
Correspondingly, the standard symplectic form on $\mathbb{F}_q^{2n}$ is defined by
$$
\left[(\mathbf{x},\mathbf{z}),(\mathbf{x}',\mathbf{z}')\right]_s=\mathbf{x}\mathbf{z}'^T-\mathbf{z}\mathbf{x}'^T.
$$

For an \(\mathbb{F}_q\)-linear subspace \(\mathcal{S} \subseteq \mathbb{F}_q^{2n}\), define its trace-symplectic dual by 
$$
S^{\perp_{\text{ts}}}=\big\{v:\operatorname{Tr}_{q/p}\big([u,v]_s\big)=0,\;\forall u\in S\big\}.
$$
Since $S$ is closed under scalar multiplication by elements of $\mathbb{F}_q$, together with the nondegeneracy of the trace pairing, we have
$$
\operatorname{Tr}_{q/p}\big(\alpha[u,v]_s\big)=0,\;\forall\alpha\in\mathbb{F}_q \iff [u,v]_s=0.
$$
Therefore,
$$
S^{\perp_{\text{ts}}}=S^{\perp_s}.
$$
This implies that commutation with all Pauli operators corresponding to labels in $S$ is equivalent to the label belonging to $S^{\perp_s}$. In the subsequent discussion, an $\mathbb{F}_q$-generating row stands for the operators corresponding to all its $\mathbb{F}_q$-multiples. Given a basis $\{\beta_1,\beta_2,\dots,\beta_m\}$ of $\mathbb{F}_q/\mathbb{F}_p$, this generating row can be realized by the operators associated with the labels $\beta_j u$.
\section{Dual-Code Galois Pairing and EA-CSS Stabilizer Realization}
\subsection{Galois Cross Gram Matrices and Galois Relative Hulls}
This section relates the Galois cross Gram matrix to the Galois relative hull and the commutation matrix of the associated CSS-type Pauli generators. These relations provide the algebraic basis for comparing different Galois indices of a fixed code pair. The classical input codes remain fixed, whereas Frobenius twisting changes the second label code and may alter the rank of the cross pairing.

Let $\mathbb{F}_q$ be a finite field of order $q = p^m$, and let $C_1,C_2 \subseteq \mathbb{F}_q^n$ be two linear codes of dimensions $r_1$ and $r_2$, with full-rank generator matrices $G_1 \in \mathbb{F}_q^{r_1\times n}$ and $G_2 \in \mathbb{F}_q^{r_2\times n}$, respectively. For any integer $0 \le \ell < m$, recall that the $\ell$-th Galois adjoint matrix is defined as $G_2^{\dagger_\ell} = \left(G_2^{[m-\ell]}\right)^T$. Using the right Galois-dual identity in \eqref{(1)}, define the $\ell$-th Galois cross Gram matrix of $(C_1,C_2)$ by
\begin{equation}
M_\ell = G_1 G_2^{\dagger_\ell} = G_1 \left(G_2^{[m-\ell]}\right)^T.
\label{(2)}
\end{equation}
If the generator matrix is replaced by \(G_i' = U_i G_i\), where \(U_i\) is invertible, then

\[
M_\ell' = U_1 M_\ell \left(U_2^{[m-\ell]}\right)^T.
\]
Therefore, \(\operatorname{rank} M_\ell\) does not depend on the choice of generator matrices, but is an invariant of the fixed ordered code pair and the index \(\ell\).

To relate this algebraic invariant to the entanglement requirement, consider CSS-type Pauli generators with $X$-type label space $\mathcal{C}_1$ and $Z$-type label space $\mathcal{C}_2^{[m-\ell]}$. The Frobenius twist realizes the Galois pairing as an ordinary cross pairing, yielding the symplectic label matrix
$$
\Gamma_\ell = \begin{pmatrix}
G_1 & 0 \\
0 & G_2^{[m-\ell]}
\end{pmatrix}.
$$
The standard symplectic form on $\mathbb{F}_q^{2n}$ is represented by the matrix
$$
J_n = \begin{pmatrix}
0 & I_n \\
-I_n & 0
\end{pmatrix}.
$$
The corresponding symplectic commutation matrix for the Pauli generators with labels given by the rows of $\Gamma_\ell$ is
$$
\Omega_\ell = \Gamma_\ell J_n \Gamma_\ell^T = \begin{pmatrix}
0 & M_\ell \\
-M_\ell^T & 0
\end{pmatrix}.
$$
Therefore,
\begin{equation}
\operatorname{rank}_{\mathbb{F}_q} \Omega_\ell = 2\operatorname{rank}_{\mathbb{F}_q} M_\ell.
\label{(3)}
\end{equation}
Thus, a symplectic decomposition of the label space generated by the rows of \(\Gamma_\ell\) contains exactly \(\operatorname{rank}_{\mathbb F_q}M_\ell\) symplectic pairs.

\begin{definition}(Galois Relative Hull)
For an ordered code pair \((C_1, C_2)\), define
\[
\mathrm{Hull}_\ell(C_1, C_2) = C_1 \cap C_2^{\perp_\ell},
\]
which is called the \(\ell\)-Galois relative hull of \(C_1\) with respect to \(C_2\), and its dimension is denoted by \(h_\ell(C_1, C_2)\).
For the exchanged code pair, consider the complementary index \(m - \ell\), and define

\begin{align*}
\operatorname{Hull}_{m-\ell}(C_2,C_1) &= C_2 \cap C_1^{\perp_{m-\ell}},\\ h_{m-\ell}(C_2,C_1) &= \dim\operatorname{Hull}_{m-\ell}(C_2,C_1).
\end{align*}

This definition aligns with the Galois relative hull in \cite{21}, adapted to the right-dual convention (the index $\ell$ here corresponds to $m-\ell \pmod{m}$ therein).

\end{definition}

The two relative hulls admit a direct interpretation through the center of the associated CSS-type Pauli subgroup. Let $S_\ell=C_1\times C_2^{[m-\ell]}.$
By \eqref{(1)}, its symplectic dual is 
\[
S_\ell^{\perp_s} = \big(C_2^{[m-\ell]}\big)^\perp \times C_1^\perp = C_2^{\perp_\ell} \times C_1^\perp.
\]
Consequently, modulo global phases, the center of the Pauli subgroup generated by \(S_\ell\) has label space
\[
\begin{aligned}
S_{\ell,0} = S_\ell \cap S_\ell^{\perp_s}= \big(C_1 \cap C_2^{\perp_\ell}\big) \times \big(C_2^{[m-\ell]} \cap C_1^\perp\big).
\end{aligned}
\]
Since
\[
C_2^{[m-\ell]} \cap C_1^\perp = (C_2 \cap C_1^{\perp_{m-\ell}})^{[m-\ell]},
\]
we obtain
\[
\dim S_{\ell,0} = h_\ell(C_1, C_2) + h_{m-\ell}(C_2, C_1).
\]
The relative hulls in the two directions correspond to the \(X\)-type and \(Z\)-type central labels that require no compensation on Bob's side, respectively. The following theorem quantifies this connection by expressing the pairing rank as the codimension of the corresponding relative hull in each input code, and establishes the reciprocity relation under exchanging the two codes and replacing the index by its complement.

\begin{theorem}\label{theorem1} (Rank Formula and Reciprocity for Galois Relative Hulls)
For an ordered pair $(C_1,C_2)$ of classical linear codes of dimensions $r_1$ and $r_2$, respectively, the $\ell$-th Galois cross Gram matrix $M_\ell$ satisfies
$$
\operatorname{rank}_{\mathbb{F}_q} M_\ell = r_1 - h_\ell(C_1,C_2) = r_2 - h_{m-\ell}(C_2,C_1).
$$
\end{theorem}

\begin{proof}
Define the linear map
\[
f_\ell : C_1 \to \mathbb{F}_q^{r_2},\quad f_\ell(x) = x\big(G_2^{[m-\ell]}\big)^T.
\]
Since the row space of $G_2$ is $C_2$, for any $x \in C_1$, the condition $f_\ell(x) = 0$ holds if and only if
\[
x\big(z^{[m-\ell]}\big)^T = 0 \qquad \text{for every } z \in C_2.
\]
By \eqref{(1)}, this is equivalent to $x\in \big(C_2^{[m-\ell]}\big)^\perp = C_2^{\perp_\ell}$. Therefore,
\[
\ker f_\ell = C_1 \cap C_2^{\perp_\ell} = \operatorname{Hull}_\ell(C_1,C_2).
\]
For every $a \in \mathbb{F}_q^{r_1}$,
\[
f_\ell(aG_1) = aG_1\big(G_2^{[m-\ell]}\big)^T = aM_\ell.
\]
Since $G_1$ has full row rank, the map $a \mapsto aG_1$ is an isomorphism onto $C_1$. Hence, by the rank-nullity theorem,
\[
\operatorname{rank}_{\mathbb{F}_q} M_\ell = \dim\operatorname{im} f_\ell = r_1 - h_\ell(C_1,C_2).
\]

On the other hand, applying the $\ell$-th Frobenius automorphism to $M_\ell$ gives $M_\ell^{[\ell]} = G_1^{[\ell]}G_2^T$. Since transposition and Frobenius twists preserve rank, the symmetric argument applied to the pair $(C_2,C_1)$ with index $m-\ell$ immediately yields $\operatorname{rank}_{\mathbb{F}_q} M_\ell = \operatorname{rank}_{\mathbb{F}_q}\big(M_\ell^{[\ell]}\big)^T = r_2 - h_{m-\ell}(C_2,C_1).$
Combining the two rank identities completes the proof.
\end{proof}

Writing $M_j(A,B)$ for the $j$-th Galois cross Gram matrix of the ordered pair $(A,B)$, the reciprocity relation takes the form
$$
\operatorname{rank}_{\mathbb{F}_q} M_\ell(C_1,C_2) = \operatorname{rank}_{\mathbb{F}_q} M_{m-\ell}(C_2,C_1).
$$

\subsection{Two-Code Galois Cross Frobenius EA-CSS Construction}
For each index \(\ell\), the ordinary EA-CSS construction applied to \((C_1, C_2^{[m-\ell]})\) realizes the cross-Galois pairing introduced above. We give an explicit Bob-side completion to determine the required entanglement, the logical dimension, and the minimum weight of nontrivial Alice-side logical operators.

The role of preshared entanglement can be illustrated by a single \(q\)-dimensional maximally entangled state
\[
|\Phi_q\rangle_{AB}
=\frac{1}{\sqrt q}\sum_{a\in\mathbb F_q}|a\rangle_A|a\rangle_B.
\]
This state satisfies
\[
[X(a)\otimes X(a)]|\Phi_q\rangle_{AB}
=|\Phi_q\rangle_{AB},
\]
and
\[
[Z(b)\otimes Z(-b)]|\Phi_q\rangle_{AB}
=|\Phi_q\rangle_{AB}.
\]
These identities explain how preshared maximally entangled pairs compensate for the noncommutativity of the symplectic pairs. Theorem 2 gives an explicit completion on Bob's side and determines the required entanglement, the logical dimension, and the exact minimum distance. Throughout the construction, errors are assumed to act only on Alice's subsystems, while Bob's subsystems are assumed noiseless. To formulate the distance statement, for any subset \(A\subseteq\mathbb F_q^n\setminus\{0\}\), define
\[
d_H(A) = \min_{\mathbf{x}\in A} \mathrm{wt}_H(\mathbf{x}),
\]
with the convention $d_H(\emptyset) = +\infty$.

 \begin{theorem}\label{theorem2}(EA-CSS Construction via Galois Pairing)
Let \(C_1, C_2 \subseteq \mathbb{F}_q^n\) be linear codes of dimensions \(r_1, r_2\), with full-row-rank generator matrices \(G_1, G_2\), respectively. For \(0 \leq \ell < m\), let
\begin{equation}
\begin{aligned}
c_\ell &= \operatorname{rank} M_\ell=
\operatorname{rank}_{\mathbb{F}_q}\left(G_1
\big(G_2^{[m-\ell]}\big)^\mathrm{T}\right),\\
K_\ell &= n - r_1 - r_2 + c_\ell.
\end{aligned}
\label{(4)}
\end{equation}
If \(K_\ell>0\), there exists an entanglement-assisted CSS code with parameters 
\[
[[n, K_\ell, d_\ell; c_\ell]]_q,
\]
whose exact minimum distance is
\begin{equation}
d_\ell = \min \left\{ d_H(C_2^{\perp_\ell} \setminus C_1), \, d_H(C_1^\perp \setminus C_2^{[m-\ell]}) \right\}.
\label{(5)}
\end{equation}
In particular, $d_\ell \ge \min\{d(C_1^\perp), d(C_2^\perp)\}$. Moreover, \(c_{\ell}\) is the minimum number of \(q\)-dimensional maximally entangled pairs required for an \(\mathbb{F}_q\)-linear stabilizer completion of \(C_1 \times C_2^{[m - \ell]}\), among all such completions that preserve the prescribed \(X\)- and \(Z\)-label spaces.
\end{theorem}

\begin{proof}
Write \(c=c_\ell\). Alice and Bob preshare \(c\) copies of \(|\Phi_q\rangle_{AB}\), with joint state
\[
|\Phi_q\rangle_{AB}^{\otimes c}
=q^{-c/2}\sum_{a\in\mathbb F_q^c}|a\rangle_A|a\rangle_B.
\]
Define \(B_\ell:=G_2^{[m-\ell]}\) and take a rank factorization
\[
M_\ell=G_1B_\ell^T=UV^T,
\qquad
U\in\mathbb F_q^{r_1\times c},
\quad
V\in\mathbb F_q^{r_2\times c},
\]
where both \(U\) and \(V\) have full column rank.
For any row vectors $a\in\mathbb{F}_q^{r_1}$, $b\in\mathbb{F}_q^{r_2}$, define operators acting on Alice's $n$ quantum subsystems and Bob's $c$ quantum subsystems by
\begin{equation*}
\begin{aligned}
\widehat{X}(a) &= X_A(a G_1) \otimes X_B(a U), \\
\widehat{Z}(b) &= Z_A(b B_\ell) \otimes Z_B(-b V).
\end{aligned}
\end{equation*}

The cross pairing of the two families of operators satisfies
$$
(aG_1)(bB_\ell)^T+(aU)(-bV)^T=a(M_\ell-UV^T)b^T=0,
$$
hence
$$
\big[\widehat{X}(a),\widehat{Z}(b)\big]=0.
$$
Thus
$$
\widehat{\mathcal{S}}=\big\{\widehat{X}(a)\widehat{Z}(b):a\in\mathbb{F}_q^{r_1},\;b\in\mathbb{F}_q^{r_2}\big\}
$$
forms an abelian stabilizer group. Owing to the full row rank of $G_1$ and $B_\ell$, this group contains no nontrivial scalar operators, and
$$
|\widehat{\mathcal{S}}|=q^{r_1+r_2}.
$$

The quantum code space is the common $+1$ eigenspace of the stabilizer operators:
\[
\mathcal{\mathcal{Q}} = \left\{ |\psi\rangle \in \mathcal{H}_q^{\otimes(n+c)} : g|\psi\rangle = |\psi\rangle,\; \forall g \in \widehat{S} \right\}.
\]

To determine the dimension of $\mathcal{\mathcal{Q}}$, define the stabilizer average
\[
\Pi_{\mathcal{\mathcal{Q}}} = \frac{1}{|\widehat{S}|}\sum_{g\in\widehat{S}} g = \frac{1}{q^{r_1+r_2}}\sum_{g\in\widehat{S}} g.
\]
The action of $\Pi_{\mathcal{\mathcal{Q}}}$ can be understood through the orthogonal decomposition into $\mathcal{\mathcal{Q}}$ and $\mathcal{\mathcal{Q}}^{\perp}$. If $|\psi\rangle\in\mathcal{\mathcal{Q}}$, every stabilizer element fixes $|\psi\rangle$, and hence $\Pi_{\mathcal{\mathcal{Q}}}|\psi\rangle = |\psi\rangle$. If $|\psi\rangle\in\mathcal{\mathcal{Q}}^{\perp}$, the identity $h\Pi_{\mathcal{\mathcal{Q}}} = \Pi_{\mathcal{\mathcal{Q}}}$ for all $h\in\widehat{S}$ shows that $\Pi_{\mathcal{\mathcal{Q}}}|\psi\rangle\in\mathcal{\mathcal{Q}}$. Meanwhile, for every $|\phi\rangle\in\mathcal{\mathcal{Q}}$,
\[
\langle\phi|\Pi_{\mathcal{\mathcal{Q}}}|\psi\rangle = \frac{1}{|\widehat{S}|}\sum_{g\in\widehat{S}}\langle g^\dagger\phi|\psi\rangle = \langle\phi|\psi\rangle = 0.
\]
Thus $\Pi_{\mathcal{\mathcal{Q}}}|\psi\rangle$ also belongs to $\mathcal{\mathcal{Q}}^{\perp}$ and must be zero. Consequently, for an arbitrary vector decomposed as
\[
|\psi\rangle = |\psi_{\mathcal{\mathcal{Q}}}\rangle + |\psi_{\perp}\rangle,
\]
we have
\[
\Pi_{\mathcal{\mathcal{Q}}}|\psi\rangle = |\psi_{\mathcal{\mathcal{Q}}}\rangle.
\]
Therefore, $\Pi_{\mathcal{\mathcal{Q}}}$ is precisely the orthogonal projector onto $\mathcal{\mathcal{Q}}$.

It follows that $\dim \mathcal{Q} = \operatorname{Tr}(\Pi_{\mathcal{Q}})$. Every nonscalar generalized Pauli operator has zero trace. As $\widehat{S}$ contains no scalar operators other than the identity, only the identity contributes to the trace sum. Consequently,
\[
\dim \mathcal{Q} = \operatorname{Tr}(\Pi_{\mathcal{Q}}) = \frac{q^{n+c}}{q^{r_1+r_2}} = q^{n+c-r_1-r_2}.
\]
Hence, by \eqref{(4)}, the code encodes $K_\ell = n - r_1 - r_2 + c_\ell$ logical qudits.

For any $\mathbb{F}_q$-linear stabilizer completion using $c'$ Bob-side subsystems, the Bob-side label matrix $\Gamma_B$ satisfies
$$
\Omega_\ell = -\Gamma_B J_{c'} \Gamma_B^T,
$$
where $J_{c'}$ is the standard symplectic matrix. By (3) and (4),
$$
2c = \operatorname{rank}_{\mathbb{F}_q} \Omega_\ell \le \operatorname{rank}_{\mathbb{F}_q} J_{c'} = 2c'.
$$
The above construction attains $c' = c$, proving that $c_\ell$ is minimal among all such completions preserving the prescribed X- and Z-label spaces.

From $S_\ell=C_1\times C_2^{[m-\ell]}$, we have
$$
S_\ell^{\perp_s}=C_2^{\perp_\ell}\times C_1^\perp.
$$
Since \(M_\ell=UV^T\), where both \(U\) and \(V\) have full column rank,
\[
aU=0\iff aM_\ell=0,\qquad
bV=0\iff bM_\ell^T=0.
\]
Consequently, the completed stabilizers acting trivially on Bob have Alice labels precisely in \(S_{\ell,0}=S_\ell\cap S_\ell^{\perp_s}\). Ignoring global phases, the label space of logical Pauli operators on Alice's side is therefore
\[
L_\ell=S_\ell^{\perp_s}/S_{\ell,0}.
\]
By Theorem 1 and \eqref{(4)},
$$
\dim_{\mathbb{F}_q}\mathcal{L_\ell}=(2n-r_1-r_2)-(r_1+r_2-2c_\ell)=2K_\ell,
$$
which matches the dimension of the Pauli label space for $K_\ell$ logical quantum subsystems.

For $v=(x,z)$, define the symplectic weight
$$
\operatorname{wt}_{s}(v)=\big|\{i:(x_i,z_i)\neq(0,0)\}\big|.
$$
When $K_\ell>0$, the minimum distance of the code equals the minimal support size of nontrivial logical Pauli operators on Alice's side, i.e.,
$$
d_\ell=\min_{v\in S_\ell^{\perp_s}\setminus S_{\ell,0}}\operatorname{wt}_{s}(v).
$$
Since
$$
\operatorname{wt}_{s}(x,z)\ge\max\big\{\operatorname{wt}_H(x),\operatorname{wt}_H(z)\big\},
$$
and every element in $S_\ell^{\perp_s}\setminus S_{\ell,0}$ has at least one component not lying in the corresponding hull, while pure $X$-type or pure $Z$-type labels can attain the corresponding minimum weight, we therefore have
$$
d_\ell=\min\Big\{d_H(C_2^{\perp_\ell}\setminus C_1),\;d_H\big(C_1^\perp\setminus C_2^{[m-\ell]}\big)\Big\}.
$$

The Frobenius automorphism preserves the Hamming weight, so
$$
d_\ell\ge\min\big\{d(C_2^{\perp_\ell}),d(C_1^\perp)\big\}=\min\big\{d(C_2^\perp),d(C_1^\perp)\big\}.
$$
\end{proof}

\begin{remark}
The set differences in (5) exclude stabilizer labels and retain nontrivial logical operators. When \(c_\ell=0\), the construction reduces to an ordinary CSS code, and (5) becomes its standard minimum-distance formula. Both set differences are nonempty whenever \(K_\ell>0\).
\end{remark}

\begin{remark}
When the two codes have systematic generator matrices under the same coordinate arrangement

\[
G_1 = [I_r \mid A], \qquad G_2 = [I_r \mid B],
\]

we have

\[
c_\ell = \operatorname{rank} \left( I_r + A (B^{[m-\ell]})^T \right).
\]

This representation requires the two codes to use common information coordinates; applying different coordinate permutations to the two codes separately will generally change the pairing rank of the original code pair.
\end{remark}

 \begin{corollary}
Let \(C_1, C_2\) both be \([n, r, n - r + 1]_q\) MDS codes, where $1 \le r \le n/2$. If
\[
K_\ell = n - 2r + c_\ell > 0,
\]
then the above construction yields an
\[
[[n, n - 2r + c_\ell, r + 1; c_\ell]]_q
\]
EAQMDS code.
\end{corollary}

\begin{proof}
The Euclidean duals of both codes are \([n, n - r, r + 1]_q\) MDS codes, so \(d_\ell \geq r + 1\). On the other hand,

\[
\dim(C_1^\perp) = n - r,
\]

while

\[
\dim(C_1^\perp \cap C_2^{[m-\ell]}) = r - c_\ell.
\]

Since \(K_\ell > 0\), the latter is a proper subspace of \(C_1^\perp\). The minimum-weight codewords of an MDS code span the entire code space: after writing its generator matrix in systematic form, every generator row has minimum weight \(r + 1\). Therefore, there exists at least one codeword of weight \(r + 1\) belonging to

\[
C_1^\perp \setminus C_2^{[m-\ell]}.
\]
The exact distance formula \eqref{(5)} gives $d_\ell \le r+1$, and hence $d_\ell = r+1$. Since $d_\ell - 1 = r \le n/2$, the entanglement-assisted quantum Singleton bound\cite[Corollary~9]{grassl2022entropic} gives
$$
K_\ell \le n - 2(d_\ell - 1) + c_\ell.
$$
 The constructed code satisfies
$$
K_\ell = n - 2r + c_\ell = n - 2(d_\ell - 1) + c_\ell,
$$
and therefore attains this bound with equality. Hence, it is an EAQMDS code.
\end{proof}

\begin{corollary}
If \(C_1 \subseteq C_2^{\perp_\ell}\), then \(c_\ell = 0\), and the construction degenerates into an ordinary CSS code. If \(C_1 \cap C_2^{\perp_\ell} = \{0\}\), then \(c_\ell = r_1\), and in this case necessarily \(r_1 \leq r_2\). In particular, when \(r_1 = r_2 = r\), the relative hulls in both directions vanish simultaneously, and \(c_\ell = r\).
\end{corollary}

\begin{corollary}(Euclidean and Hermitian Specializations)
Under the hypotheses of Theorem 2, the construction specializes to the Euclidean and Hermitian cases as follows.

1. At $\ell = 0$, Theorem 2 recovers the Euclidean two-code EA-CSS construction of Galindo et al. \cite[Th.~4]{bib11}, with entanglement cost
$$
c_0 = \operatorname{rank}_{\mathbb{F}_q}\big(G_1 G_2^T\big) = r_1 - \dim_{\mathbb{F}_q}\big(C_1 \cap C_2^\perp\big).
$$

2. If $m$ is even, the specialization at $\ell = m/2$ uses the Hermitian pairing over $\mathbb{F}_q$. Applying the Euclidean two-code construction \cite[Th.~4]{bib11} to $(C_1, C_2^{[m/2]})$ and using the relative-hull identity in \cite[eq. (1)]{21} gives
$$
c_{m/2} = \operatorname{rank}_{\mathbb{F}_q}\left(G_1 \big(G_2^{[m/2]}\big)^T\right) = r_1 - \dim_{\mathbb{F}_q}\big(C_1 \cap C_2^{\perp_\mathrm{H}}\big),
$$
where $C_2^{\perp_\mathrm{H}}$ denotes the Hermitian dual of $C_2$.

\end{corollary}

\begin{corollary}(Single-Code Specialization)
For \(C_1=C_2=C\) in Theorem 2, write \(r=\dim_{\mathbb F_q}C\) and \(h_\ell=\dim_{\mathbb F_q}(C\cap C^{\perp_\ell})\). If \(n-r-h_\ell>0\), there exists an EA-CSS code with parameters
\[
[[n,n-r-h_\ell,d_\ell;r-h_\ell]]_q,
\]
where \(d_\ell\geq d(C^\perp)\). 
These parameters follow directly from Theorem 2 and agree with those obtained by Liu et al. \cite[Cor. 2.5]{liu2019new} using $k$-Galois duality.
\end{corollary}

\subsection{Galois Index Selection for Fixed Classical Code Pairs}
Theorem \ref{theorem2} expresses the entanglement cost and the number of encoded logical qudits in terms of the cross-Galois pairing rank, while providing a distance lower bound independent of the index. We therefore study Galois index selection with the ordered code pair $(C_1,C_2)$ and its coordinate arrangement fixed. Define the entanglement consumption sequence and its attainable value set by
$$
\boldsymbol{c}(C_1,C_2) = (c_0,c_1,\dots,c_{m-1}), \qquad \mathcal{E}(C_1,C_2) = \{c_\ell : \ell \in \mathbb{Z}_m\}.
$$

For each index $\ell$ with $K_\ell>0$, the associated EA-CSS code satisfies
\begin{align*}
K_\ell &= n - r_1 - r_2 + c_\ell,\\ d_\ell &\ge d_* := \min\{d(C_1^\perp),d(C_2^\perp)\}.
\end{align*}

For general code pairs, the exact minimum distance in (5) may depend on the Galois index, whereas the lower bound \(d_*\) is independent of the index. The following proposition bounds \(c_\ell\) and describes how the entanglement cost and the number of encoded logical qudits vary together as the index changes.
\begin{proposition}\label{proposition1}(Entanglement Consumption and Logical Dimension)
Under the above conditions,
$$
\max\{0, r_1+r_2-n\} \le c_\ell \le \min\{r_1,r_2\},
$$
and
\begin{equation}
K_\ell - c_\ell = n - r_1 - r_2.
\label{(6)}
\end{equation}
For any $\ell,\ell'\in\mathbb{Z}_m$, we have
$$
c_{\ell'} - c_\ell = h_\ell(C_1,C_2) - h_{\ell'}(C_1,C_2),
$$
and
$$
K_{\ell'} - K_\ell = c_{\ell'} - c_\ell.
$$
\end{proposition}

 \begin{proof}
By the rank formula for the relative hull,
$$
c_\ell = r_1 - h_\ell(C_1,C_2).
$$
Since
$$
0 \le h_\ell(C_1,C_2) \le \min\{r_1, n-r_2\},
$$
we obtain $c_\ell \ge \max\{0, r_1+r_2-n\}$. The matrix $M_\ell$ is of size $r_1\times r_2$, so $c_\ell \le \min\{r_1,r_2\}$.
The remaining identities follow from Theorem 1 and \eqref{(4)}.

\end{proof}
Proposition \ref{proposition1} shows that index variation changes the entanglement cost and the number of encoded logical qudits by the same amount, while preserving the net encoding dimension. Resource selection must therefore account for the required number of encoded logical qudits.

\begin{definition}(Frobenius Period)
For a linear code $C \subseteq \mathbb{F}_q^n$, define its Frobenius period by
$$
a(C) := \min\{a \in \mathbb{Z}_{>0} : C^{[a]} = C\}.
$$
Since $C^{[m]} = C$, this period is well defined and satisfies $a(C) \mid m$.
\end{definition}
The Frobenius periods of the input codes yield the following index reduction.
\begin{proposition}\label{p2}(Index Reduction via Frobenius Periods)
Let $a_i = a(C_i)$ for $i = 1,2$, and set
$$
g = \gcd(a_1,a_2).
$$
Then, for every $\ell \in \mathbb{Z}_m$,
$$
c_{\ell+g} = c_\ell,\qquad K_{\ell+g} = K_\ell.
$$
If $K_\ell > 0$, then
$$
d_{\ell+g} = d_\ell.
$$
Consequently, examining $\ell = 0,\dots,g-1$ suffices to cover the parameter configurations arising from all Galois indices.
\end{proposition}

\begin{proof}
Let $S_\ell = C_1 \times C_2^{[m-\ell]},$ with all Frobenius indices interpreted modulo $m$. By the definition of $a_i = a(C_i)$,
\[
C_i^{[a_i]} = C_i, \qquad i = 1,2.
\]
The invertibility of Frobenius twisting also gives $C_i^{[-a_i]} = C_i$.
For the shift $a_2$, we immediately have
\[
S_{\ell+a_2} = C_1 \times C_2^{[m-\ell-a_2]} = C_1 \times C_2^{[m-\ell]} = S_\ell.
\]
Hence the corresponding quantum parameters coincide.

For the shift $a_1$, observe that
\[
\big(S_{\ell+a_1}\big)^{[a_1]} = C_1^{[a_1]} \times C_2^{[m-\ell]} = S_\ell.
\]
We show that this Frobenius correspondence preserves the quantum parameters. Let $\Gamma$ be a full-row-rank generator matrix of $S_{\ell+a_1}$. Then $\Gamma^{[a_1]}$ generates $S_\ell$. Since the entries of $J_n$ lie in the prime field,
\[
\Gamma^{[a_1]} J_n \big(\Gamma^{[a_1]}\big)^T = \big(\Gamma J_n \Gamma^T\big)^{[a_1]}.
\]
Frobenius twisting preserves matrix rank, so the symplectic commutation matrices associated with these two label spaces have the same rank. Equations \eqref{(3)} and \eqref{(4)} therefore give $c_{\ell+a_1}=c_\ell$ and
$K_{\ell+a_1}=K_\ell$.

To compare the minimum distances, suppose that $K_\ell > 0$. For any $x,y \in \mathbb{F}_q^{2n}$,
\[
\bigl[x^{[a_1]}, y^{[a_1]}\bigr]_s = \bigl[x,y\bigr]_s^{p^{a_1}}.
\]
Together with the bijectivity of Frobenius twisting, this identity implies
\[
\big(S_{\ell+a_1}^{\perp_s}\big)^{[a_1]} = S_\ell^{\perp_s}.
\]
Moreover, writing $S_{j,0} = S_j \cap S_j^{\perp_s}$, we have
\[
S_{\ell+a_1,0}^{[a_1]} = \big(S_{\ell+a_1} \cap S_{\ell+a_1}^{\perp_s}\big)^{[a_1]} = S_{\ell,0}.
\]
Thus Frobenius twisting induces a bijection
\[
S_{\ell+a_1}^{\perp_s} \setminus S_{\ell+a_1,0} \longrightarrow S_\ell^{\perp_s} \setminus S_{\ell,0}.
\]
This bijection preserves symplectic weight because it preserves the zero pattern of every coordinate pair. The proof of Theorem 2 identifies the distance in \eqref{(5)} with the minimum symplectic weight on the corresponding set difference. The weight-preserving bijection therefore gives $d_{\ell+a_1}=d_\ell$.

We have therefore established, for $i = 1,2$,
\[
c_{\ell+a_i} = c_\ell,\qquad K_{\ell+a_i} = K_\ell,
\]
and $d_{\ell+a_i} = d_\ell$ whenever $K_\ell > 0$. Since these identities hold for every $\ell \in \mathbb{Z}_m$, they also imply invariance under the negative shifts $-a_1$ and $-a_2$. Now let $g = \gcd(a_1,a_2)$. By B\'ezout's identity, there exist integers $u,v$ such that
\[
g = u a_1 + v a_2.
\]
Repeated application of the preceding identities yields
\[
c_{\ell+g} = c_\ell,\qquad K_{\ell+g} = K_\ell.
\]
If $K_\ell > 0$, every intermediate shift preserves this positivity, so the distance identities likewise give
\[
d_{\ell+g} = d_\ell.
\]
Finally, $a_1,a_2\mid m$ implies $g\mid m$. Hence every index is congruent modulo $g$ to an element of $\{0,1,\dots,g-1\}$, and examining these $g$ indices suffices. 

\end{proof}

In particular, if either input code satisfies $C_i^{[1]} = C_i$, then $g = 1$, so all indices yield the same quantum parameters. Thus effective index selection depends on the Frobenius structure and relative orthogonality of the fixed code pair.

By \eqref{(6)}, a reduction in the entanglement cost is accompanied by an equal reduction in the number of encoded logical qudits. Index selection therefore concerns the tradeoff between these resources under the common distance lower bound.

When $n - r_1 - r_2 > 0$, all indices yield positive logical dimensions, and minimizing the entanglement cost is equivalent to maximizing the relative hull dimension.
Accordingly, define
$$
c_{\min}=\min_{\ell\in\mathbb Z_m}c_\ell=r_1-\max_{\ell\in\mathbb Z_m}h_\ell(C_1,C_2),
$$
and the optimal index set
$$
\mathcal L_{\min}=\mathop{\arg\min}_{\ell\in\mathbb Z_m}c_\ell.
$$
By Proposition \ref{p2}, it suffices to search the indices $0,1,\dots,g-1$; all other minimizing indices follow by shifts that are multiples of $g$.

To encode at least $K_0>0$ logical qudits, minimize $c_\ell$ subject to
$$
c_\ell \ge K_0 - n + r_1 + r_2.
$$
Conversely, for an entanglement budget $C_{\text{bud}}$, maximize $K_\ell$ subject to
$$
c_\ell \le C_{\text{bud}}.
$$
These optimizations are over the attainable set $\mathcal{E}(C_1,C_2)$, which may have gaps; feasibility therefore depends on the actual attainable values.

To quantify the entanglement savings relative to the conventional Euclidean and Hermitian choices, define
\begin{equation}
\Delta_\text{E} = c_0 - c_\text{min},\qquad \Delta_\text{H} = c_{m/2} - c_\text{min},
\label{(7)}
\end{equation}
where the second quantity is defined only when $m$ is even.

A positive $\Delta_\text{E}$ or $\Delta_\text{H}$ indicates that the corresponding conventional index does not minimize the entanglement requirement of the fixed code pair.

\section{Index Tuning and EAQMDS Construction Based on Fixed GRS Code Pairs} 
The index selection framework is now applied to fixed GRS code pairs. Evaluation over the multiplicative group of the field expresses the entries of the Galois cross Gram matrix as finite field power sums, yielding an explicit modular intersection formula for the entanglement requirement. This formula characterizes its dependence on the Galois index while keeping the input code pair fixed. Conditions for the resulting EA-CSS codes to attain the entanglement-assisted Singleton bound are then established, leading to EAQMDS constructions.

\subsection{Fixed GRS Code Pairs and Their Entanglement Consumption}
 
Let $q = p^m$ with $m>1$, and put $n = q-1$. Fix an ordering $(a_1,a_2,\dots,a_n)$ of $\mathbb{F}_q^*$ and integers $r,t$ satisfying
$$
1 \le r < \frac{n}{2},\qquad 0 \le t < n.
$$
Define
$$
G_1 = \big(a_v^i\big)_{\substack{0\le i<r,\\ 1\le v\le n}},\qquad
G_2 = \big(a_v^{t+j}\big)_{\substack{0\le j<r,\\ 1\le v\le n}},
$$
and let
$$
C_1 = \operatorname{row}(G_1),\qquad C_2 = \operatorname{row}(G_2).
$$
These are GRS codes with parameters $[n,r,n-r+1]_q$, sharing the same evaluation points and having multiplier vectors $(1,1,\dots,1)$ and $(a_1^t,a_2^t,\dots,a_n^t)$, respectively.

Throughout this subsection, $(r,t)$ and the evaluation-point ordering remain fixed, while only the Galois index $\ell\in\mathbb{Z}_m$ varies. Since $2r<n$, the corresponding logical dimension satisfies
$$
K_\ell = n - 2r + c_\ell > 0
$$
for every index. To express $c_\ell$ through modular intersections, write
$$
I_r = \{0,1,\dots,r-1\}\subseteq\mathbb{Z}_n,\qquad s_\ell = p^{m-\ell}.
$$

For this code pair, the finite-field power-sum identity shows that the nonzero entries of $M_\ell$ are determined by the congruence $i + s_\ell(t + j) \equiv 0 \pmod{n}$. The resulting matrix structure converts the rank computation into the modular-intersection count stated below.

\begin{theorem}\label{t3}(Entanglement Formula for Fixed GRS Code Pairs)
For the fixed GRS code pair defined above and every $0\le \ell < m$, the entanglement consumption is
\begin{equation}
c_\ell(t,r) = \big|I_r \cap \big(-s_\ell(t + I_r)\big)\big|,
\label{(8)}
\end{equation}
where translations and scalar multiplication are taken in $\mathbb{Z}_n$. The corresponding EA-CSS code has parameters
$$
\big[\big[n,\; n-2r + c_\ell(t,r),\; r+1;\; c_\ell(t,r)\big]\big]_q
$$
and is EAQMDS.
\end{theorem}
\begin{proof}
By \eqref{(2)}, the $(i,j)$-entry of $M_\ell$ is given by the following power sum for $0 \le i,j < r$:
\begin{align*}
(M_\ell)_{ij} &= \sum_{a\in\mathbb{F}_q^*} a^{i+s_\ell(t+j)}\\
&=
\begin{cases}
-1,  i + s_\ell(t + j) \equiv 0 \pmod{n},\\
0,  \text{otherwise},
\end{cases}
\end{align*}
by the power sum identity over $\mathbb{F}_q^*$.

Since $\gcd(s_\ell,n) = 1$, the map $j \mapsto -s_\ell(t + j)$ is injective modulo $n$. Hence each row and each column of $M_\ell$ contains at most one nonzero entry, so its rank equals the number of nonzero entries. Therefore, by \eqref{(4)}, the entanglement cost $c_\ell(t,r)$ equals the number of nonzero entries of $M_\ell$,
$$
\begin{aligned}
c_\ell(t,r) &= \operatorname{rank}_{\mathbb{F}_q} M_\ell \\
&= \#\{j \in I_r : -s_\ell(t + j) \bmod n \in I_r\} \\
&= \big|I_r \cap \big(-s_\ell(t + I_r)\big)\big|.
\end{aligned}
$$
Since both input codes are $[n,r,n-r+1]_q$ MDS codes and $K_\ell = n-2r + c_\ell(t,r) > 0$, Corollary 1 yields the stated EAQMDS parameters.
\end{proof}

For fixed $r, t$, varying the Galois index changes $c_\ell$ and $K_\ell$ by the same amount, while preserving the code length $n$, exact minimum distance $r+1$, and net encoding dimension $K_\ell - c_\ell = n - 2r$. Thus index selection provides entanglement tuning at a fixed exact distance, with every resulting code remaining EAQMDS.

\subsection{A Family of Parameter Sets with Closed Form Formulas for Entanglement Savings}
For the fixed shift $t = n - r$, the modular intersection count in \eqref{(8)} admits a closed form evaluation under the condition stated below. The resulting expression quantifies the entanglement savings relative to the Euclidean index while preserving the exact minimum distance and EAQMDS property.

\begin{theorem}\label{t4}(Closed Form Entanglement Cost for a Fixed GRS Code Pair)
Let $q = p^m$, $m \ge 3$, and $n = q - 1$. Fix
$$
3 \le r < \frac{n}{2},\qquad t = n - r,
$$
and let $(C_1,C_2)$ be the GRS code pair defined in Section IV-A. Then
$$
c_0 = r - 1.
$$
For every $1 \le \ell < m$ satisfying $p^\ell r \le n$,
$$
c_\ell = \left\lfloor \frac{r}{p^\ell} \right\rfloor,\qquad c_0 - c_\ell = r - 1 - \left\lfloor \frac{r}{p^\ell} \right\rfloor > 0.
$$
The corresponding EAQMDS code has parameters
$$
\left[\!\left[n,\; n - 2r + \left\lfloor \frac{r}{p^\ell} \right\rfloor,\; r + 1;\; \left\lfloor \frac{r}{p^\ell} \right\rfloor \right]\!\right]_q.
$$
\end{theorem}
\begin{proof}
By \eqref{(8)} and $p^\ell s_\ell \equiv 1 \pmod{n}$, the entanglement consumption $c_\ell$ equals the number of solutions to
$$
p^\ell i + j \equiv r \pmod{n},\qquad 0\le i,j<r.
$$
For $\ell=0$, the inequality $2r<n$ reduces this congruence to $i+j=r$, whose solutions are $(i,r-i)$ for $1\le i\le r-1$. Hence $c_0 = r-1$.

For $1\le \ell<m$, put $h = p^\ell$ and assume $hr \le n$. Then
$$
0 \le hi + j \le (h+1)(r-1) < n + r.
$$
The only integer in this range congruent to $r$ modulo $n$ is $r$ itself. Thus the admissible solutions are exactly
$$
i = 1,2,\dots,\left\lfloor \frac{r}{h} \right\rfloor,\qquad j = r - hi,
$$
giving
$$
c_\ell = \left\lfloor \frac{r}{p^\ell} \right\rfloor.
$$
Since $r \ge 3$ and $p^\ell \ge 2$, this is strictly smaller than $c_0 = r - 1$, proving the stated saving. Substituting this value of $c_\ell$ into the parameter expression in Theorem \ref{t3} gives the stated EAQMDS parameters.

\end{proof}
Theorem \ref{t4} gives a nonincreasing entanglement cost over the indices satisfying $1 \le \ell < m$ and $p^\ell r \le n$. Outside this range, modular wraparound may introduce additional solutions, and the general counting formula in Theorem \ref{t3} applies.

In particular, if $pr \le n$, choosing $\ell = 1$ gives
$$
c_0 - c_1 = r - 1 - \left\lfloor \frac{r}{p} \right\rfloor > 0.
$$
Since $m \ge 3$, this index differs from the Euclidean index and, when applicable, the Hermitian index. Thus index selection strictly reduces the entanglement consumption relative to the Euclidean realization while preserving the exact minimum distance and EAQMDS property.

By Theorem \ref{t4}, $r < p^\ell$ gives $c_\ell = 0$ under the stated hypotheses. The resulting code is an ordinary CSS MDS code with the stated parameters.

\begin{corollary}(Zero-Entanglement Realization under Fixed Classical Inputs)
Under the hypotheses of Theorem \ref{t4}, if an index $1 \le \ell < m$ satisfies
$$
r < p^\ell,\qquad p^\ell r \le n,
$$
then $c_\ell = c_{\min} = 0$, and the fixed code pair yields an ordinary CSS MDS code with parameters
$$
\left[\!\left[n,\, n - 2r,\, r + 1\right]\!\right]_q.
$$
\end{corollary}
The following family illustrates the reduction from positive entanglement consumption to zero while keeping the classical inputs fixed. Let $q = 2^{m},\, m \ge 5$, and choose
$$
n = 2^{m} - 1,\qquad r = 3,\qquad t = n - 3.
$$
By Theorem \ref{t4}, the indices $\ell = 0,1,2$ give
$$
(c_0,c_1,c_2) = (2,1,0),
$$
respectively, yielding
$$
\left[\!\left[n,\, n-4,\,4;\,2\right]\!\right]_{2^{m}},\quad
\left[\!\left[n,\, n-5,\,4;\,1\right]\!\right]_{2^{m}},\quad
\left[\!\left[n,\, n-6,\,4\right]\!\right]_{2^{m}}.
$$
All three codes have exact minimum distance $4$ and net encoding dimension $n-6$. Since \(m\ge5\), the index \(\ell=2\) differs from the Euclidean index and, when \(m\) is even, from the Hermitian index.

\subsection{Examples and Comparisons}
The following example compares the entanglement requirements of Euclidean, Hermitian, and general Galois realizations arising from the same fixed GRS code pair. Take
$$
q = 16,\qquad n = 15,\qquad r = 6,\qquad t = 9.
$$
Both input codes have parameters $[15,6,10]_{16}$. Since $m=4$, the Euclidean and Hermitian choices correspond to $\ell=0$ and $\ell=2$, respectively, while $\ell=1,3$ are general Galois indices. Applying the modular intersection formula in Theorem \ref{t3} gives the entanglement costs and EAQMDS parameters listed in Table \ref{Table 1}.

\begin{table*}[htbp]
  \centering
  \caption{Entanglement consumption and EAQMDS parameters for the fixed GRS pair over \(\mathbb{F}_{16}\) with \((r, t) = (6, 9)\).}
\label{Table 1}
  \begin{tabular}{ccccc}
    \toprule
    $\ell$ & Pairing Type & $I_6 \cap \big[-s_\ell(9+I_6)\big]$ & $c_\ell$ & Quantum Code Parameters \\
    \midrule
    $0$ & Euclidean & $\{1,2,3,4,5\}$ & $5$ & $[[15,8,7;5]]_{16}$ \\
    $1$ & General Galois & $\{1,2,3\}$ & $3$ & $[[15,6,7;3]]_{16}$ \\
    $2$ & Hermitian & $\{1,4,5\}$ & $3$ & $[[15,6,7;3]]_{16}$ \\
    $3$ & General Galois & $\{2,4\}$ & $2$ & $[[15,5,7;2]]_{16}$ \\
    \bottomrule
  \end{tabular}
\end{table*}

Table \ref{Table 1} shows that the minimum entanglement consumption is $c_{\min}=2$, attained at the general Galois index $\ell=3$, with savings
$$
\Delta_{\text{E}}=3,\qquad \Delta_{\text{H}}=1.
$$
Thus the minimum for this fixed code pair is attained outside the Euclidean and Hermitian choices. All four realizations remain EAQMDS with exact minimum distance $7$ and net encoding dimension $K_\ell - c_\ell = 3$; the reduction in entanglement consumption is accompanied by an equal reduction in logical dimension.

Further examples over different finite fields are summarized in Table \ref{2}. For each row, the field, evaluation points, dimension $r$, and shift $t$ remain fixed, while only the Galois index varies. The complete entanglement sequences are obtained from Theorem \ref{t3}, and every index yields an EAQMDS realization with exact minimum distance $r+1$.

\begin{table*}
\centering
\caption{Complete entanglement sequences and minimum-entanglement EAQMDS realizations for fixed GRS code pairs.}
\label{2}
\scriptsize
\begin{tabular}{c c c c c}
\toprule
\(q\) & \((r, t)\) & Complete sequence \((c_0,c_1, \dots, c_{m-1})\) & Optimal index & Corresponding quantum code \\
\midrule
64 & \((12, 49)\) & \((9, 6, 3, 2, 0, 3)\) & 4 & \([[63, 39, 13; 0]]_{64}\) \\
64 & \((13, 48)\) & \((10, 6, 3, 2, 1, 4)\) & 4 & \([[63, 38, 14; 1]]_{64}\) \\
256 & \((48, 197)\) & \((37, 24, 12, 12, 9, 6, 1, 13)\) & 6 & \([[255, 160, 49; 1]]_{256}\) \\
\(81 = 3^4\) & \((15, 62)\) & \((11, 5, 3, 1)\) & 3 & \([[80, 51, 16; 1]]_{81}\) \\
\bottomrule
\end{tabular}
\label{tab:grs_eaqmds}
\end{table*}

The first two code pairs over \(\mathbb{F}_{64}\) illustrate realizations with zero and low entanglement consumption, respectively. For $(r,t)=(12,49)$, the general Galois index $\ell=4$ reduces the entanglement consumption from the Euclidean and Hermitian values $9$ and $2$ to zero, yielding the ordinary CSS MDS code $[[63,39,13]]_{64}$. For $(r,t)=(13,48)$, the same index requires only one entangled pair, compared with $10$ and $2$ for the Euclidean and Hermitian choices, respectively.

The example over $\mathbb{F}_{81}$ demonstrates the same effect in odd characteristic. For $(r,t)=(15,62)$, the Euclidean and Hermitian entanglement costs are $11$ and $3$, whereas the general Galois index $\ell=3$ requires only one pair and yields an EAQMDS code with parameters $[[80,51,16;1]]_{81}$.

The fixed GRS code pair over $\mathbb{F}_{256}$, with $(r,t)=(48,197)$, illustrates the extent to which Galois index selection can reduce entanglement consumption while preserving the exact minimum distance. Its Euclidean, Hermitian, and minimum entanglement costs are
$$
c_0 = 37,\qquad c_4 = 9,\qquad c_6 = c_{\min} = 1.
$$
At the minimizing index $\ell=6$,
$$
I_{48} \cap \big[-4(197 + I_{48})\big] = \{44\},
$$
which certifies the pairing rank $c_6=1$. Hence
$$
\Delta_{\text{E}} = 36,\qquad \Delta_{\text{H}} = 8.
$$

A logical dimension constraint can change the preferred index. For this last code pair,
$K_\ell = 159 + c_\ell$, so the requirement $K_\ell \ge 165$ is equivalent to $c_\ell \ge 6$. The complete sequence in Table \ref{2} shows that the minimum feasible cost is $c_5 = 6$, yielding
$$
[[255, 165, 49; 6]]_{256}.
$$
Both the Euclidean and Hermitian choices satisfy the same dimension constraint but require $37$ and $9$ entangled pairs, respectively. Thus constrained index selection still saves $31$ and $3$ pairs relative to these choices.

Reference \cite{21} established the existence of EAQMDS codes with the corresponding parameters through monomial equivalence transformations. This section keeps the GRS code pair and its coordinate ordering fixed and adjusts entanglement consumption solely through the choice of the Galois index. The presented code families and examples demonstrate that general Galois indices can yield lower entanglement consumption than the Euclidean or Hermitian choices for the same code pair while preserving the exact minimum distance. These results provide an explicit approach to realizing these known parameters and optimizing entanglement consumption without replacing the constituent classical codes.
\section{Index Selection and Entanglement Control for Fixed BCH Code Pairs}
For fixed cyclic code pairs, Galois index selection can be described through their defining sets. Frobenius twisting induces modular multiplication on these sets, allowing the pairing rank to be computed by a defining-set intersection or, equivalently, a weighted overlap of cyclotomic cosets. Specialization to BCH codes yields exact entanglement costs and logical dimensions, together with a minimum distance lower bound independent of the index.

\subsection{Defining-Set Representation and Index Action}

Let $q = p^m$ and $\gcd(n,p) = 1$, and write $\mathbb{Z}_n = \mathbb{Z}/n\mathbb{Z}$. Fix a primitive $n$-th root of unity $\alpha$ in an extension of $\mathbb{F}_q$. For $a\in\mathbb{Z}_n$, define the $q$-cyclotomic coset
$$
Q_q(a) = \{a q^j \bmod n : j\ge 0\}.
$$

For a union $T\subseteq\mathbb{Z}_n$ of $q$-cyclotomic cosets, let $B(T)$ denote the cyclic code with defining set $T$ relative to $\alpha$. Since $x^n-1$ has no repeated roots,
$$
\dim B(T) = n - |T|.
$$

Write $T^c = \mathbb{Z}_n\setminus T$, and let $Z(C)$ denote the defining set of a cyclic code $C$. Euclidean duality, Frobenius twisting, and code intersection satisfy
$$
\begin{aligned}
Z\big(B(T)^\perp\big) &= -T^c,\\
Z\big(B(T)^{[u]}\big) &= p^u T,\qquad 0\le u < m,\\
Z\big(B(T_1)\cap B(T_2)\big) &= T_1\cup T_2,
\end{aligned}
$$
where set arithmetic is taken modulo $n$. Frobenius twisting is semilinear over $\mathbb{F}_q$ and preserves dimension and Hamming weight.

Fix two defining sets $T_1,T_2$, each a union of $q$-cyclotomic cosets, and let
$$
B_i = B(T_i),\qquad i=1,2,
$$
have parameters $[n,k_i,D_i]_q$. Set
$$
C_i = B_i^\perp,\qquad \dim C_i = n-k_i.
$$
Throughout the following construction, $T_1,T_2$ remain fixed, and only the Galois index $\ell\in\mathbb{Z}_m$ varies. This choice allows the entanglement cost to be computed from the defining sets, while Theorem 2 gives the common distance lower bound
$$
d_\ell \ge \min\{D_1,D_2\}
$$
whenever the logical dimension is positive.

\subsection{Entanglement Consumption Formula for Fixed Cyclic Code Pairs}
The defining-set identities above turn the rank formula in Theorem 1 into an explicit intersection count, yielding the following entanglement formula and EA-CSS parameters.

\begin{theorem}\label{t5}(Entanglement Formula for Fixed Cyclic Code Pairs)
For the fixed cyclic code pair defined above and every $0 \le \ell < m$, the entanglement consumption is
\begin{equation}
c_\ell = \big|T_1 \cap (-p^{m-\ell}T_2)\big| = \big|T_2 \cap (-p^\ell T_1)\big|.
\label{(10)}
\end{equation}
The corresponding Galois relative hull dimension is
$$
h_\ell(C_1,C_2) = \big|T_1 \setminus (-p^{m-\ell}T_2)\big| = |T_1| - c_\ell.
$$
If
$$
K_\ell = n - |T_1| - |T_2| + c_\ell > 0,
$$
then there exists an entanglement-assisted CSS code with parameters
$$
[[n,K_\ell,d_\ell;c_\ell]]_q,\qquad d_\ell \ge \min\{D_1,D_2\}.
$$
\end{theorem}
\begin{proof}
Since $C_i = B_i^\perp$, the right Galois-dual convention gives
$$
C_2^{\perp_\ell} = (C_2^\perp)^{[m-\ell]} = B_2^{[m-\ell]}.
$$
The defining-set identities therefore yield
$$
\begin{aligned}
h_\ell(C_1,C_2) &= n - \big|(-T_1^c) \cup p^{m-\ell}T_2\big| \\
&= \big|T_1 \setminus (-p^{m-\ell}T_2)\big|.
\end{aligned}
$$
Since $r_1 = |T_1|$, Theorem 1 and the definition of $c_\ell$ give
$$
c_\ell = |T_1| - h_\ell(C_1,C_2) = \big|T_1 \cap (-p^{m-\ell}T_2)\big|.
$$
Multiplication by $-p^\ell$ permutes $\mathbb{Z}_n$, and $qT_2 = T_2$ because $T_2$ is a union of $q$-cyclotomic cosets. Hence
$$
\begin{aligned}
c_\ell &= \big|(-p^\ell T_1) \cap qT_2\big| \\
&= \big|T_2 \cap (-p^\ell T_1)\big|.
\end{aligned}
$$
By Theorem 2, the logical dimension is $$K_\ell = n - |T_1| - |T_2| + c_\ell.$$ If $K_\ell > 0$, the resulting EA-CSS code satisfies $d_\ell \ge \min\{D_1, D_2\}$.

\end{proof}
Theorem \ref{t5} reduces index selection to a finite overlap minimization problem over indices with positive logical dimension. Define
$$
\mathcal{I}_+ = \{\ell \in \mathbb{Z}_m : K_\ell > 0\}.
$$
Whenever $\mathcal{I}_+ \neq \varnothing$, an entanglement minimizing admissible index satisfies
$$
\ell_{\min} \in \arg\min_{\ell\in\mathcal{I}_+} \big|T_1 \cap \big(-p^{m-\ell}T_2\big)\big|.
$$
Thus, the entanglement cost can be optimized directly from the fixed defining sets while ensuring a positive logical dimension.

Since
$$
K_\ell - c_\ell = n - |T_1| - |T_2|,
$$
reducing the entanglement cost decreases the number of encoded logical qudits by the same amount. To encode at least \(K_0>0\) logical qudits, the minimization is restricted to indices satisfying
$$
c_\ell \ge K_0 - n + |T_1| + |T_2|.
$$
The overlap can also be evaluated at the level of cyclotomic cosets. Let $\mathcal{Q}_i$ be the collection of distinct $q$-cyclotomic cosets constituting $T_i$, and define
$$
\mu_\ell\big(Q_q(a)\big) = Q_q(-p^{m-\ell}a).
$$
Because the multiplier is coprime to $n$, this map permutes the $q$-cyclotomic cosets and preserves their sizes. Consequently,
$$
c_\ell = \sum_{Q\in\mathcal{Q}_1\cap\mu_\ell(\mathcal{Q}_2)} |Q|.
$$
The entanglement cost is therefore a weighted overlap of coset collections, with each common coset contributing its cardinality. Counting common cosets alone is insufficient when their sizes differ.

Finally, Frobenius symmetries of the defining sets can reduce the index search. In particular, if $p^a T_2 = T_2$, then
$$
c_{\ell+a}=c_\ell,\qquad K_{\ell+a}=K_\ell,
$$
with indices taken modulo $m$. Hence indices related by this symmetry need only be evaluated once when optimizing the entanglement cost.

These formulas specialize directly to fixed BCH code pairs. Their defining sets are generated by consecutive exponents, allowing the entanglement cost to be evaluated by the preceding intersection formula, while the BCH bound supplies a common minimum distance guarantee.

For a starting position \(b\) and a designed distance \(\delta \geq 2\), define
\[
T(b, \delta) = \bigcup_{j=0}^{\delta-2} Q_q(b + j),
\]
and denote
\[
B(q, n, b, \delta) = B(T(b, \delta)).
\]

Since the defining set contains consecutive indices
\[
b, b+1, \dots, b+\delta-2,
\]
the BCH bound guarantees
\[
d(B(q, n, b, \delta)) \geq \delta.
\]
Fix two BCH codes
\[
B_i = B(q, n, b_i, \delta_i), \qquad i = 1, 2,
\]
and their defining sets \(T_i = T(b_i, \delta_i)\). By Theorem \ref{t5}, every index satisfying
\[
K_\ell = n - |T_1| - |T_2| + c_\ell > 0
\]
yields an EA-CSS code with parameters
\[
[[n, K_\ell, d_\ell; c_\ell]]_q, \qquad d_\ell \geq \min\{\delta_1, \delta_2\}.
\]

With $b_1,b_2,\delta_1,\delta_2$ fixed, index selection varies the entanglement cost and the number of encoded logical qudits while preserving the BCH distance guarantee.

\subsection{Attainable Entanglement Levels for a Fixed BCH Code Pair over \(\mathbb{F}_8\)}

The following example illustrates that the attainable entanglement levels may have gaps and that the entanglement cost need not vary monotonically with the Galois index. Fix
$$q = 8,\quad n = 63,\quad b_1 = 1,\quad b_2 = 37,\quad \delta_1 = \delta_2 = 10,$$
and consider the BCH code pair
$$B_1 = B(8,63,1,10),\quad B_2 = B(8,63,37,10).$$
Their defining sets are
$$T_1 = \{1,2,3,4,5,6,7,8,9,16,24,32,40,48,56\},$$
$$T_2 = \{5,13,21,29,37,38,39,40,41,42,43,44,45,52,60\}.$$
Since $|T_1| = |T_2| = 15$, both codes have dimension 48 and minimum distance at least 10.

The entanglement costs and logical dimensions in Tables III-V are obtained from Theorem \ref{t5}. The corresponding minimum distance guarantees follow from the BCH bound.

\begin{table}[h]
\centering
\caption{Galois index comparison for the fixed BCH pair over \(\mathbb{F}_8\) with \(n = 63\), \((b_1, b_2) = (1, 37)\), and \(\delta_1 = \delta_2 = 10\).}
\label{3}
\begin{tabular}{c c c c c}
\hline
\(\ell\) & \(T_1 \cap (-2^{3-\ell}T_2)\) & \(c_\ell\) & \(K_\ell\) & Quantum code parameters \\
\hline
0 & \(\{3, 24\}\) & 2 & 35 & \([[63, 35, d_0; 2]]_8\), \(d_0 \geq 10\) \\
1 & \(\{9\}\) & 1 & 34 & \([[63, 34, d_1; 1]]_8\), \(d_1 \geq 10\) \\
2 & \(\{5, 6, 40, 48\}\) & 4 & 37 & \([[63, 37, d_2; 4]]_8\), \(d_2 \geq 10\) \\
\hline
\end{tabular}
\end{table}

The entanglement consumption sequence is
$$(c_0,c_1,c_2)=(2,1,4),$$
so the minimum $c_{\text{min}}=1$ is attained at the general Galois index $\ell=1$. Compared with the Euclidean realization, this choice saves one maximally entangled pair and reduces the number of encoded logical qudits from 35 to 34. All three realizations retain the distance guarantee $d_\ell\ge 10$ and the net encoding dimension $K_\ell-c_\ell=33$.

Moreover, the attainable entanglement set is
$$\mathcal{E}(C_1,C_2)=\{1,2,4\}.$$
The missing intermediate value $3$ shows that index selection need not realize every integer between the minimum and maximum entanglement costs. The decrease from $c_0$ to $c_1$, followed by the increase to $c_2$, also demonstrates that the entanglement cost need not vary monotonically with the index.

\subsection{Entanglement Savings for Fixed BCH Code Pairs over $\mathbb{F}_{16}$}

The following two examples compare Euclidean, Hermitian, and general Galois realizations for fixed BCH code pairs over $\mathbb{F}_{16}$. Both examples have length $n = 255$, with designed distances 12 and 16, respectively. Since $m = 4$, the Euclidean and Hermitian indices are $\ell = 0$ and $\ell = 2$, while $\ell = 1,3$ are general Galois indices. Write $Q(a) = Q_{16}(a)$.

For designed distance 12, consider
$$B_1 = B(16,255,1,12), \quad B_2 = B(16,255,235,12),$$
with defining sets
$$T_1 = \bigcup_{a=1}^{11} Q(a), \quad T_2 = \bigcup_{a=235}^{245} Q(a).$$
Here $|T_1| = 22$ and $|T_2| = 21$, so the two BCH codes have dimensions 233 and 234, respectively, and minimum distances at least 12.

For designed distance 16, consider instead
$$B_1 = B(16,255,1,16), \quad B_2 = B(16,255,232,16),$$
with defining sets
$$T_1 = \bigcup_{a=1}^{15} Q(a), \quad T_2 = \bigcup_{a=232}^{246} Q(a).$$
In this case, $|T_1| = 30$ and $|T_2| = 29$, giving dimensions 225 and 226, respectively, and minimum distances at least 16.

For each pair, the classical inputs remain fixed while $\ell$ varies. Theorem \ref{t5} gives the intersections and resource parameters in Table \ref{4}. Each row yields an EA-CSS code
$$[[255, K_\ell, d_\ell; c_\ell]]_{16},\qquad d_\ell \ge \delta,$$
where $\delta$ denotes the common designed distance of the corresponding input pair.
\begin{table}[h]
\centering
\caption{Galois index comparison for two fixed BCH code pairs over \(\mathbb F_{16}\) of length 255.}
\label{4}
\begin{tabular}{c c c c c}
\hline
$\delta$ & $\ell$ & $T_1 \cap (-2^{4-\ell}T_2)$ & $c_\ell$ & $K_\ell$ \\
\hline
12 & 0 & $Q(1) \cup Q(10) \cup Q(11)$ & 6 & 218 \\
12 & 1 & $\displaystyle\bigcup_{a=5}^{10} Q(a)$ & 12 & 224 \\
12 & 2 & $Q(3) \cup Q(4) \cup Q(5)$ & 6 & 218 \\
12 & 3 & $Q(2)$ & 2 & 214 \\
16 & 0 & $\displaystyle Q(1) \cup \bigcup_{a=9}^{15} Q(a)$ & 16 & 212 \\
16 & 1 & $\displaystyle\bigcup_{a=5}^{11} Q(a)$ & 14 & 210 \\
16 & 2 & $Q(3) \cup Q(4) \cup Q(5)$ & 6 & 202 \\
16 & 3 & $Q(2)$ & 2 & 198 \\
\hline
\end{tabular}
\end{table}

For the pair with designed distance 12, the entanglement consumption sequence is
\[
(c_0,c_1,c_2,c_3) = (6,12,6,2).
\]
The minimum is attained at $\ell = 3$, saving four maximally entangled pairs relative to either conventional index. The resulting code has parameters
\[
[[255,214,d_3;2]]_{16},\qquad d_3 \ge 12,
\]
and all four realizations satisfy $K_\ell - c_\ell = 212$.

For the pair with designed distance 16, the sequence becomes
\[
(c_0,c_1,c_2,c_3) = (16,14,6,2).
\]
Again, $\ell = 3$ attains the minimum, giving
\[
[[255,198,d_3;2]]_{16},\qquad d_3 \ge 16.
\]
The savings relative to the Euclidean and Hermitian realizations are 14 and 4, respectively, while $K_\ell - c_\ell = 196$ for every index.

Thus, both fixed pairs admit a general Galois realization requiring only two entangled pairs. The second pair provides a larger distance guarantee, with a lower net encoding dimension. Within each pair, reducing entanglement consumption also reduces the number of encoded logical qudits by the same amount. Consequently, the preferred index depends on the required number of encoded logical qudits: for example, if $K_\ell \ge 218$ is imposed on the pair with designed distance 12, the minimum feasible entanglement cost is 6, attained at the Euclidean and Hermitian indices.

\subsection{General Galois Index Selection in Long BCH Codes}

To illustrate index selection at a larger code length, take
\[
\begin{aligned}
q &= 64,\quad n = 4095,\quad b_1 = 1, \\ b_2& = 4077,
\delta_1 = \delta_2 = 16,
\end{aligned}
\]
and fix the BCH code pair
\[
B_1 = B(64,4095,1,16),\qquad B_2 = B(64,4095,4077,16).
\]
Their defining sets are
\[
T_1 = \bigcup_{a=1}^{15} Q_{64}(a),\qquad T_2 = \bigcup_{a=4077}^{4091} Q_{64}(a).
\]
Since $|T_1| = |T_2| = 30$, both input codes have dimension $4065$ and minimum distance at least $16$.

Write $Q(a) = Q_{64}(a)$. Applying Theorem \ref{t5} gives the defining-set intersections and logical dimensions listed in Table \ref{5}. Every coset appearing in these intersections has size two, so each common coset contributes two to the entanglement cost. Consequently,
\[
(c_0,c_1,c_2,c_3,c_4,c_5) = (24,16,8,4,2,8).
\]

\begin{table}[h]
\centering
\caption{Defining-set intersections and quantum dimensions for the fixed BCH pair over \(\mathbb{F}_{64}\) with \(n = 4095\), \((b_1, b_2) = (1, 4077)\), and \(\delta_1 = \delta_2 = 16\).}
\label{5}
\begin{tabular}{c c c c}
\hline
\(\ell\) & \(T_1 \cap (-2^{6-\ell}T_2)\) & \(c_\ell\) & \(K_\ell\) \\
\hline
0 & \(\bigcup_{a=4}^{15} Q(a)\) & 24 & 4059 \\
1 & \(\bigcup_{a=2}^{9} Q(a)\) & 16 & 4051 \\
2 & \(Q(1) \cup Q(2) \cup Q(3) \cup Q(4)\) & 8 & 4043 \\
3 & \(Q(1) \cup Q(2)\) & 4 & 4039 \\
4 & \(Q(1)\) & 2 & 4037 \\
5 & \(Q(8) \cup Q(10) \cup Q(12) \cup Q(14)\) & 8 & 4043 \\
\hline
\end{tabular}
\end{table}
The Euclidean and Hermitian indices correspond to $\ell=0$ and $\ell=3$, with entanglement costs $24$ and $4$, respectively. The minimum $c_{\text{min}} = 2$ is attained at the general Galois index $\ell=4$, yielding
\[
[[4095,4037,d_4;2]]_{64},\qquad d_4 \ge 16.
\]
The corresponding entanglement savings are
\[
\Delta_{\text{E}} = 22,\qquad \Delta_{\text{H}} = 2.
\]

All six realizations retain the distance guarantee $d_\ell \ge 16$ and satisfy
\[
K_\ell - c_\ell = 4035.
\]
Thus, the reduction in entanglement consumption is accompanied by an equal reduction in the number of encoded logical qudits. The subsequent increase from $c_4 = 2$ to $c_5 = 8$ further shows that the entanglement cost is not monotone in the Galois index.

Table \ref{6} summarizes the minimum entanglement costs and the savings relative to the Euclidean and Hermitian realizations for the three BCH examples over fields of even extension degree.

\begin{table*}[tbp]
\centering
\caption{Entanglement savings of optimal Galois indices relative to the Euclidean and Hermitian indices for fixed BCH code pairs.}
\label{6}
\setlength{\tabcolsep}{5pt}
\footnotesize
\begin{tabular}{c c c c c c}
\hline
Fixed BCH code pair $(q,n,b_1,b_2,\delta)$ & $c_0$ & $c_{m/2}$ & $c_{\min}$ & Optimal index & $(\Delta_\mathrm{E},\Delta_\mathrm{H})$ \\
\hline
$(16,255,1,235,12)$ & 6 & 6 & 2 & 3 & $(4,4)$ \\
$(16,255,1,232,16)$ & 16 & 6 & 2 & 3 & $(14,4)$ \\
$(64,4095,1,4077,16)$ & 24 & 4 & 2 & 4 & $(22,2)$ \\
\hline
\end{tabular}
\label{tab:galois_saving}
\end{table*}

For every code pair in Table \ref{6},
\[
\Delta_{\text{E}} > 0,\qquad \Delta_{\text{H}} > 0.
\]
Thus neither the Euclidean nor the Hermitian index minimizes entanglement consumption for these fixed inputs. 

The BCH examples show that entanglement consumption can vary nonmonotonically with the Galois index and that the set of attainable entanglement costs may contain gaps. The defining-set intersection formula determines the entanglement cost and logical dimension exactly for each index, providing an explicit method for computing the minimum entanglement cost subject to a prescribed requirement on the logical dimension and identifying the indices that attain it.
\section{Conclusion}
This paper presents an \(\mathbb F_q\)-linear EA-CSS stabilizer construction that realizes the Galois pairing of a prescribed ordered pair of classical linear codes at an arbitrary index. An explicit Bob-side completion determines the number of encoded logical qudits and yields an exact minimum distance expression. Among \(\mathbb F_q\)-linear completions preserving the prescribed label spaces, the minimum entanglement cost equals the Galois cross Gram rank and is equivalently characterized by the Galois relative hull dimension. With the original code pair and its coordinate arrangement fixed, the resulting framework characterizes the entanglement costs attainable through index selection and the corresponding numbers of encoded logical qudits. Frobenius periods can reduce the index search, while logical-qudit requirements and entanglement budgets determine the admissible indices.

For the constructed GRS code pairs, the pairing rank is expressed as a modular intersection count. The MDS structure yields EAQMDS realizations whose exact minimum distance remains unchanged across all indices. Under a specified shift, a closed form entanglement formula further identifies conditions for reduced or zero entanglement consumption. Explicit examples show that neither the Euclidean nor the Hermitian index necessarily attains the minimum. These comparisons quantify the resource changes induced solely by index selection, with the classical inputs, code length, exact minimum distance, and net encoding dimension kept unchanged.

For cyclic and BCH code pairs, the defining-set representation converts the entanglement calculation into an intersection count under Frobenius multiplication, equivalently a weighted overlap of cyclotomic cosets. For BCH inputs, this determines the exact entanglement costs and logical dimensions while retaining a common designed distance guarantee. The examples exhibit lower entanglement realizations beyond the conventional indices, nonmonotone index dependence, and gaps in the attainable entanglement levels. Consequently, resource feasibility depends on the attainable set, rather than only on its minimum and maximum values.

For fixed classical inputs, reducing entanglement consumption decreases the number of encoded logical qudits by the same amount. Index selection therefore balances entanglement availability against the required number of encoded logical qudits under a common minimum distance lower bound. Further questions include characterizing the attainable entanglement set and the least period of the entanglement consumption sequence, identifying optimal indices without exhaustive evaluation, and determining how the exact quantum minimum distance varies with the index for BCH code pairs.

\section*{Acknowledgments}
This work was supported by the Fundamental Research Funds for the Central Universities (26CX03010A), the Key Project of Computing Power Internet and Information Security, Ministry of Education (2024ZD006), China Education Innovation Research Fund (2022BL027), Shandong Provincial Natural Science Foundation of China (ZR2023LLZ013), and the Taishan Scholars Program (No. tsqn202312231). 

Mengqing Yang and Tongjiang Yan are with the College of Science, China University of Petroleum (East China), Qingdao, China (e-mail: 18456110120@126.com; yantoji@163.com). (Corresponding author: Tongjiang Yan.)


%

\bibliographystyle{IEEEtran}
\bibliography{TIT-bibliography}

\vspace{11pt}

\vspace{-33pt}
\begin{IEEEbiographynophoto}{Mengqing Yang}
is a Ph.D. student with the College of Science, China University of Petroleum (East China), Qingdao, China. Her research interests include quantum coding and entanglement-assisted quantum error-correcting codes.
\end{IEEEbiographynophoto}
\begin{IEEEbiographynophoto}{Tongjiang Yan}
was born in 1973 in Shandong Province, China. He graduated from the Department of Mathematics, Huaibei Coal Industry Teachers College, China, in 1996. He received the M.S. degree in mathematics from Northeast Normal University in 1999 and the Ph.D. degree from Xidian University in 2007. He is currently a professor with China University of Petroleum (East China). His research interests include cryptography and coding theory.
\end{IEEEbiographynophoto}

\vfill

\end{document}